\documentclass[letterpaper, 10 pt, conference]{ieeeconf}  
\usepackage{gen_settings}

\IEEEoverridecommandlockouts                              

\title{\LARGE \bf
A Barrier-Based Scenario Approach to Verify Safety-Critical Systems
}

\author{Prithvi Akella, Aaron D. Ames$^{1}$
\thanks{*This work was supported by AFOSR}
\thanks{$^{1}$All authors are with the California Institute of Technology
        {\tt\small pakella@caltech.edu}}%
}

\begin{document}

\maketitle
\thispagestyle{empty}
\pagestyle{empty}

\begin{abstract}

In this letter, we detail our randomized approach to safety-critical system verification.  Our method requires limited system data to make a strong verification statement.  Specifically, our method first randomly samples initial conditions and parameters for a controlled, continuous-time system and records the ensuing state trajectory at discrete intervals.  Then, we evaluate these states under a candidate barrier function $h$ to determine the constraints for a randomized linear program.  The solution to this program then provides either a probabilistic verification statement or a counterexample.  To show the validity of our results, we verify the robotarium simulator and identify counterexamples for its hardware counterpart.  We also provide numerical evidence to validate our verification statements in the same setting.  Furthermore, we show that our method is system-independent by performing the same verification method on a quadrupedal system in a multi-agent setting as well.

\end{abstract}

\section{Introduction}
It is natural to question the validity of controllers for safety-critical systems insofar as safety is of critical importance for these systems. Therefore, there has been a tremendous amount of work in the controls literature concerning both the development and verification of these controllers. On the developmental side, some work aims at learning or modifying existing control theoretic techniques, \textit{e.g.} control barrier and Lyapunov functions, to iteratively develop better controllers that satisfy the desired safety objectives by default~\cite{taylor2020learning, robey2020learning, boffi2020learning, khansari2014learning, ravanbakhsh2019learning, ravanbakhsh2016robust}. On a related note, there has also been interest in developing controllers against formal system specifications to ensure satisfactory operation as well~\cite{wongpiromsarn2010receding,raman2014model,kloetzer2008fully,lindemann2018control,lindemann2019control}. For the sake of completeness, we have mentioned these works, although this paper will focus more on verification.

As controller verification typically does not restrict the verification analysis to a single control form, there are multiple ways this problem has been approached. One vein of work attempts to determine a Lyapunov or barrier function for the controlled system, to act as a certificate of system stability/safety~\cite{giesl2015review, johansen2000computation,anderson2015advances,bobiti2016sampling,prajna2004safety,prajna2007framework, boffi2020learning}. Due to their exploitation of existing control techniques to simplify the verification problem, works in this vein tend to be less sample complex than works in the next paradigm. This second paradigm expresses the verification question as an optimization problem whose solution corresponds to a counterexample or a (probabilistic) verification statement~\cite{annpureddy2011s, donze2010breach, dreossi2019verifai, ghosh2018verifying, corso2020survey, akella2022test}.  These works typically associate satisfactory behavior to positive evaluations of a robustness measure over system trajectories and aim to minimize this measure over a set of parameters of interest.

Each paradigm has its benefits and shortcomings. In the former case, the reduction in sample complexity arises through \textit{apriori} knowledge of the system dynamics and controller. Except for~\cite{prajna2004safety}, these dynamics tend to only be functions of the system state and do not take into account extraneous parameters of interest, \textit{e.g.} user-defined control objectives, human input through parameterization, \textit{etc}. Additionally, \textit{apriori} knowledge of the controller may not always be provided, especially if the system's controller is a complex, layered controller, \textit{e.g.} the controller for an autonomous car, a quadruped, a bipedal exoskeleton, \textit{etc}. The latter case permits more flexibility in this vein. Specifically, they only require the capacity to quantify system satisfaction of its objective through a robustness measure with positive robustness indicating objective satisfaction. Then, these methods try to determine the minimum robustness over a given set of parameters. However, they do not make as efficient use of existing control techniques due to their black-box system assumptions. As a result, these optimization problems suffer from poor performance in higher dimensions. Therefore, we aim to address both these shortcomings through our work at the intersection of these paradigms.

\begin{figure}[t]
    \centering
    \includegraphics[width = 0.49\textwidth]{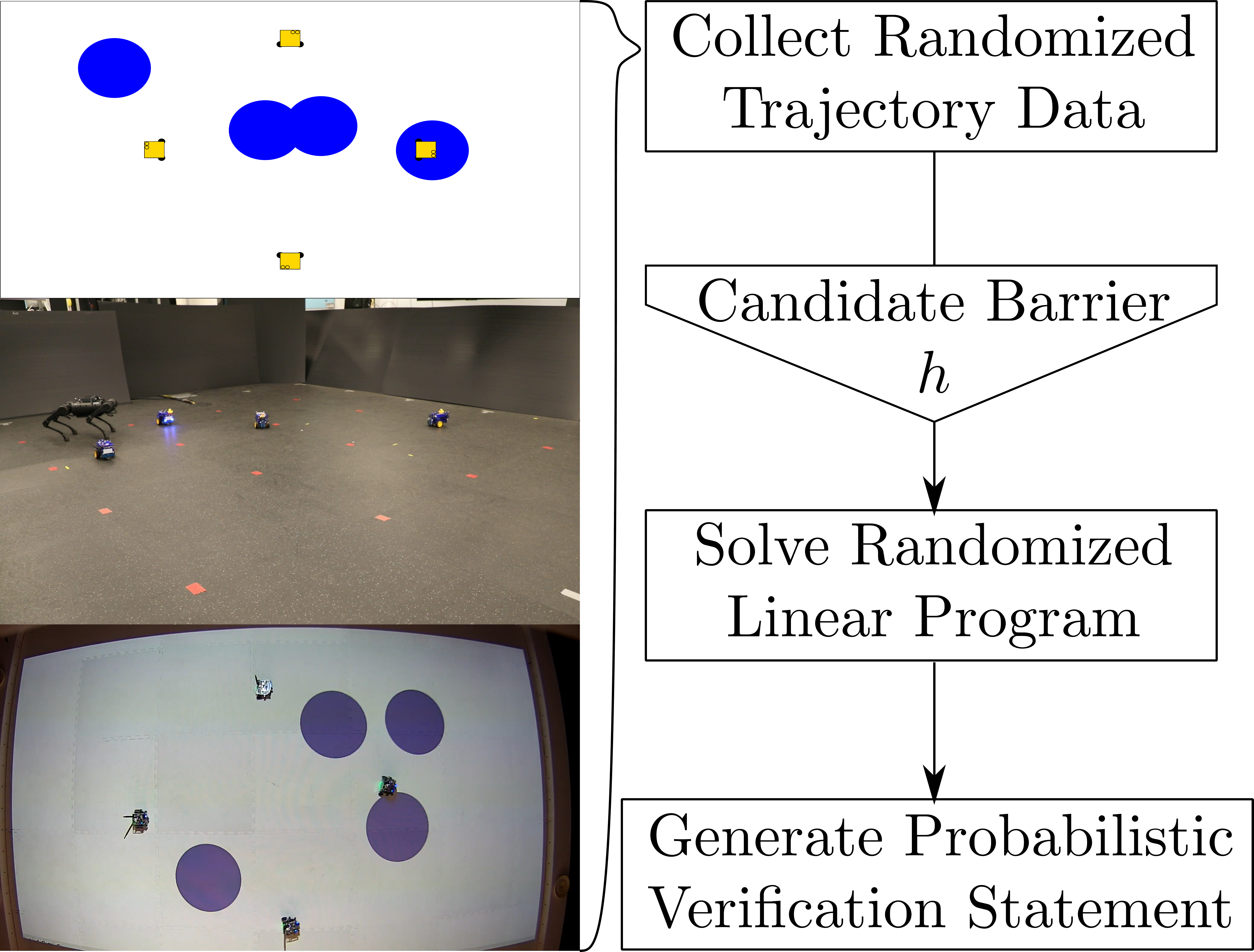}
    \caption{An overview of the approach detailed within the paper.  By collecting randomized state trajectory data of the system-to-be-verified and evaluating all transitions under a candidate barrier function $h$, we determine the constraints for a randomized linear program.  Positivity of the solution to this program corresponds to a probabilistic verification statement regarding this system's ability to render positive the candidate barrier function $h$ over its trajectories.} \vspace{-0.25 in}
    \label{fig:title}
\end{figure}

\newidea{Our Contribution:}  We aim to utilize the benefits of both paradigms to address the shortcomings of the other.  Our approach will focus on those safety-critical systems whose controllers vary with respect to a parameterized input, \textit{e.g.} varying goal locations, obstacle locations, different control objectives, \textit{etc}. For these systems, we will provide either a counterexample or probabilistic verification guarantee. More specifically, our approach will use control barrier functions evaluated over randomly sampled system trajectories to inform the constraints for a randomized linear program. The solution to this program will identify a counterexample or provide for that probabilistic guarantee. Finally, we show the efficacy of validating our verification statements for a system simulator and its hardware counterpart.  To preface our contribution, however, we will first introduce some necessary background information in the next section.

\section{Problem Formulation}
\label{sec:problem_formulation}
This section will be split into two parts.  The first part will detail some mathematical background information - discrete control barrier functions and scenario optimization.  The second part will formally state the problem under study.  To preface both parts, however, we will define some notation.

\spacing
\newidea{Notation:} $\mathbb{Z}_+$ is the set of all positive integers, and $\mathbb{R}_+ = \{x \in \mathbb{R}~|~x\geq0\}$. $|A|$ is the cardinality of the set $A$.

\subsection{Mathematical Preliminaries}
\label{sec:mathematical_preliminaries}

\newidea{Discrete Control Barrier Functions:} First introduced by Ames et al. in~\cite{ames2016control} and built upon by Agrawal et al. in~\cite{agrawal2017discrete}, discrete control barrier functions are a novel control tool designed to enforce forward invariance of their $0$-superlevel sets for nonlinear discrete-time systems such as:
\begin{equation}
    \label{eq:discrete_system}
    x_{k+1} = f(x_k,u_k),~x \in \mathcal{X}\subset\mathbb{R}^n,~u \in \mathcal{U}\subset \mathbb{R}^m.
\end{equation}

Then, a discrete exponential control barrier function (DE-CBF) $h: \mathbb{R}^n \to \mathbb{R}$ is designed to classify those control inputs that maintain positivity of the same function in a specific manner.  More accurately, if we define the $0$-superlevel set of a candidate DE-CBF $h$ as
\begin{equation}
    \label{eq:superlevel}
    \mathcal{C} = \{x \in \mathcal{X}~|~h(x) \geq 0\},
\end{equation}
then the definition of a DE-CBF is as follows:
\begin{definition}[Adapted from Definition 4 in~\cite{agrawal2017discrete}]
\label{def:discrete_control_barrier}
A function $h: \mathbb{R}^n \to \mathbb{R}$ is a \textit{discrete exponential control barrier function} if the following inequality holds for some $\gamma \in [0,1)$:
\begin{equation}
    \forall~x_k \in \mathcal{C},~\exists~u \in \mathcal{U} \suchthat h\left(f(x_k,u_k)\right) \geq \gamma h(x_k).
\end{equation}
\end{definition}
\noindent Then if we define the set of inputs that satisfy the CBF inequality in Definition~\ref{def:discrete_control_barrier} as
\begin{equation}
    K(x) = \{u \in \mathcal{U}~|~h(f(x,u)) \geq \gamma h(x)\},
\end{equation}
we have the following Theorem regarding forward invariance of the $0$-superlevel set $\mathcal{C}$ of the DE-CBF $h$.
\begin{theorem}
[Adapted from Proposition 3 in~\cite{agrawal2017discrete}]
\label{thm:disc_cbf}
For a discrete exponential control barrier function $h:\mathbb{R}^n \to \mathbb{R}$ and it's $0$-superlevel set $\mathcal{C}$, if $x_0 \in \mathcal{C}$ and all inputs $u_k \in K(x_k)~\forall~k\in\mathbb{Z}_+$, then $x_k \in \mathcal{C}~\forall~k \in \mathbb{Z}_+$.
\end{theorem}

In what will follow, we will only reference discrete exponential control barrier functions and as such will simply refer to such functions as control barrier functions.  Then, the overarching idea as to how we will utilize these functions for verification is to construct a linear program whose constraints are the inequalities mentioned in Definition~\ref{def:discrete_control_barrier}.  The decision variable will be $\gamma$ and the constraints will be randomly sampled from robot trajectories.  While solving such an optimization problem will prove rather easy, guaranteeing that the solution has meaning over all trajectories is the subject of scenario optimization which will be detailed next.

\newidea{Scenario Optimization:}  The brief description of scenario optimization in this section will stem primarily from the work done by Campi and Garrati in~\cite{campi2008exact,campi2018wait}.   Scenario optimization tries to identify robust solutions to uncertain convex optimization problems of the following form:
\begin{equation}
    \label{eq:uncertain_program}
    \tag{UP}
    \begin{aligned}
        z^* & = \argmin_{z \in \mathbb{Z} \subset \mathbb{R}^d}~ & &c^Tz, \\
        &~~\mathrm{subject~to}~ & &z \in \mathbb{Z}_{\delta},~\delta \in \Delta.
    \end{aligned}
\end{equation}
Here, \eqref{eq:uncertain_program} is the uncertain program as $\delta \in \Delta$ is an uncertain parameter with probability measure $\mathbb{P}$.  Convexity is assured via assumed convexity in the spaces $\mathbb{Z}$ and $\mathbb{Z}_{\delta}$, and typically, $|\Delta| = \infty$.  Hence, direct identification of a robust solution $z^*$ such that $z^* \in \mathbb{Z}_{\delta}~\forall~\delta \in \Delta$ is usually infeasible.

To resolve this issue, the study of scenario optimization solves a related optimization problem formed from an $N$-sized sample of the constraints $\delta$ and provides a probabilistic guarantee on the robustness of the corresponding solution $z^*_N$.  Specifically, if we were to take an $N$-sized sample of $\delta$, $\deltaset$ - termed scenarios in the scenario optimization literature - we could construct the following scenario program:
\begin{equation}
    \label{eq:scenario_program}   
    \tag{RP-N}
    \begin{aligned}
        z^*_N & = \argmin_{z \in \mathbb{Z} \subset \mathbb{R}^d}~ & &c^Tz, \\
        &~~\mathrm{subject~to}~ & &z \in \bigcap\limits_{i=1,2,\dots,N} \mathbb{Z}_{\delta_i}.
    \end{aligned}
\end{equation}
\noindent Then, we require the following assumption.
\begin{assumption}
\label{assump:RPN_solvability}
The scenario program~\eqref{eq:scenario_program} is solvable for any $N$-sample set $\deltaset$ and has a unique solution $z^*_N$.
\end{assumption}

For more information on why this assumption is made, we direct the reader to~\cite{campi2008exact,campi2018wait}.  Assumption~\ref{assump:RPN_solvability} then guarantees existence of a scenario solution $z^*_N$ for~\eqref{eq:scenario_program}.  As such, we can define a set containing those constraints $\delta \in \Delta$ to which the scenario solution $z^*_N$ is not robust, \textit{i.e.}
\begin{equation}
\label{eq:non_robust_constraint}
F(z) = \{\delta \in \Delta~|~z \not \in \mathbb{Z}_{\delta}\}.
\end{equation}
With this set definition we can formally define the \textit{violation probability} of our solution.
\begin{definition}
\label{def:violation}
The \textit{violation probability} $V(z)$ of a given $z \in \mathbb{Z}$ is defined as the probability of sampling a constraint $\delta$ to which $z$ is not robust, \textit{i.e.} $V(z) = \prob[\delta \in F(z)]$ .
\end{definition}
\noindent Then, the main theorem is as follows:
\begin{theorem}[Adapted from Theorem 1 in~\cite{campi2008exact}]
\label{thm:scenario_opt}
Let Assumption~\ref{assump:RPN_solvability} hold.  The following inequality is true:
\begin{equation}
    \prob^N[V(z^*_N) > \epsilon] \leq \sum_{i=0}^{d-1} \binom{N}{i} \epsilon^i(1-\epsilon)^{N-i}.
\end{equation}
\end{theorem}
\noindent In Theorem~\ref{thm:scenario_opt} above, $N$ is the number of sampled constraints $\delta$ for the scenario program~\eqref{eq:scenario_program}; $z^*_N$ is the scenario solution to the corresponding scenario program; $V(z^*_N)$ is the violation probability of that solution as per Definition~\ref{def:violation}; $d$ is the dimension in which $z$ lies, \textit{i.e. } $z \in \mathbb{R}^d$; and $\prob^N$ is the induced probability measure over sets of $N$-samples of $\delta$ given the probability measure $\prob$ for $\delta$.  For more information on the intersection of scenario optimization and control more generally, we direct the readers to~\cite{calafiore2012robust,calafiore2013stochastic} and the citations within.  With this information, we can now formally state the problem under study in the next section.

\subsection{Problem Statement}
We wish to verify safety properties for a given system without knowledge of the controller and/or dynamics.  As such, we will assume our system is a continuous control system whose dynamics $f$ and controller $U$ are unknown, though $U$ may depend on extraneous parameters $\theta$.
\begin{equation}
    \label{eq:nominal_system}
    \begin{gathered}
    \dot x = f(x,u),~x \in \mathcal{X} \subset \mathbb{R}^n,~u \in \mathcal{U} \subset \mathbb{R}^m, \\
    u = U(x,\theta),~\theta \in \Theta \subset \mathbb{R}^p, \\
    \dot x(t,\theta) = f\left(x(t,\theta),U(x(t,\theta),\theta) \right),~(x_0,\theta) \in \mathcal{X} \times \Theta.
    \end{gathered}
\end{equation}
As is common, $\mathcal{X}$ is the state space, $\mathcal{U}$ is the feasible control space, and $\Theta$ is the feasible parameter space.  Furthermore, $x(t,\theta)$ corresponds to the solution to the closed-loop system at time $t$ given the initial condition and parameter $(x_0,\theta)$.  Notice that we do not allow the parameter $\theta$ to vary over the system's trajectory.  Once chosen, the parameter $\theta$ is fixed.

To quantify our system's safety objective then, we will assume the capacity to measure system safety through evaluations of a candidate barrier function $h$ at specific time instances $t_k = k \Delta t$ where $k\in \mathbb{Z}_+$ and $\Delta t > 0$.
\begin{assumption}
\label{assump:barrier_existence}
We have a barrier function $h:\mathcal{X} \times \Theta \to [-m,M],~m,M \in \mathbb{R}_+$ with $0$-superlevel set $\mathcal{C} = \{(x,\theta) \in \mathcal{X} \times \Theta~|~h(x,\theta) \geq 0\}$.  Furthermore, the system's safety objective is satisfied by ensuring continued positivity of $h$ at time intervals $t_k$, \textit{i.e.} by ensuring $h(x(t_k,\theta), \theta) \geq 0,~\forall~k \in \mathbb{Z}_+$ and $(x_0,\theta) \in \mathcal{C}$.
\end{assumption}
\noindent For context, this assumption is not too restrictive.  Consider a simple example where the system is to avoid static obstacles that can be placed anywhere in a confined, 2-dimensional space.  Then, defining $h$ to be a function of the system state $x$ and static obstacle location $\theta$ whose $0$-superlevel set does not coincide with the obstacle suffices.  Also note that we are not stating that the system objective is to ensure continued positivity of $h(x,\theta)$.  Rather, we only assume that the system's objective is satisfied if $h(x,\theta)$ is kept positive, which is a significantly more relaxed assumption.  In the latter case, for example, over-approximations of obstacle regions are valid.  The former case would require specific knowledge of the obstacle locations which is significantly harder.

With these definitions and assumptions, our formal problem statement will follow.
\begin{problem}
For the closed-loop system~\eqref{eq:nominal_system} and barrier function $h$ as per Assumption~\ref{assump:barrier_existence}, devise a method to verify whether $h(x(t_k,\theta),\theta) \geq 0~\forall~k \leq K \in\mathbb{Z}_+$ and $(x_0,\theta) \in \mathcal{C}$.
\end{problem}
\noindent Ideally we would like to guarantee safety for all time indeces $k\in\mathbb{Z}_+$.  However, as we intend to use a sampling approach, we require some finite time $K$ by which to stop taking samples of the system trajectory.  However, $K$ can be any large positive integer.  With our problem formally stated, we will now move describe our approach.

\section{Main Contributions}
This section will be split into three parts.  The first part will outline the overarching idea behind our approach; the second part will state and prove our main result, two lemmas, and a corollary; and the third part will provide numerical examples indicating that we can sample from a distribution integral to our approach.  With that, our overarching idea will follow.

\spacing
\newidea{Overarching Idea:}  As stated, we want to verify whether a given safety-critical system ensures continued positivity of a candidate barrier function $h$ at specific time instances $t_k$ as per Assumption~\ref{assump:barrier_existence}.  To make this verification statement, if we could identify a safety decay rate $\gamma \in [0,1)$ satisfying the following inequality for this candidate barrier function $h$:
\begin{equation}
    \label{eq:desired_inequality}
    h(x(t_1,\theta),\theta) \geq \gamma h(x_0,\theta),~(x_0,\theta) \in \mathcal{C},
\end{equation}
then we could directly use Theorem~\ref{thm:disc_cbf} to make our desired verification statement.  However, checking the veracity of the inequality in~\eqref{eq:desired_inequality} would require checking this condition for every possible trajectory emanating from initial conditions $(x_0,\theta) \in \mathcal{C}$. With there being a (potentially) infinite amount of them, this is infeasible.  

However, we note that we can rephrase the identification of $\gamma$ into an optimization problem with infinite constraints:
\begin{align}
        \gamma^* & = \argmax_{\gamma \in \mathbb{R}}~ & &\gamma, \label{eq:actual_opt} \tag {BASE-OPT} \\
        &~~\mathrm{subject~to}~ & & h(x(t_1,\theta),\theta) \geq \gamma h(x_0,\theta), \\
        & & & \qquad \qquad \forall~(x_0,\theta) \in \mathcal{C}.
\end{align}
If the solution $\gamma^*$ to~\eqref{eq:actual_opt} were positive, then the inequality in~\eqref{eq:desired_inequality} would be true.  To relax the number of constraints for~\eqref{eq:actual_opt} and yield a solvable optimization problem, we will instead randomly sample the constraints and generate a scenario program.

This transformation of~\eqref{eq:actual_opt} to an uncertain program for which we can guarantee a scenario solution is the crux of our approach.  In doing so, we will randomly sample over feasible robot trajectories and measure system safety throughout.  Every measurement will provide a new constraint to the scenario program.  Then, a positive scenario solution $\gamma^*_N$ to the corresponding scenario program constitutes successful maintenance of the inequality~\eqref{eq:desired_inequality} with high probability.  A negative solution will identify a counterexample.  With this overarching idea in mind, we will now be more specific in our statement of our main contributions.

\subsection{Main Results}
As mentioned, our approach involves transforming~\eqref{eq:actual_opt} to an uncertain program which we will solve.  In keeping with the notation for scenario optimization utilized earlier, our uncertain program is as follows:
\begin{align}
        \gamma^* & = \argmax_{\gamma \in \mathbb{R}}~ & &\gamma, \label{eq:uncertain_opt} \tag {BASE-UP} \\
        &~~\mathrm{subject~to}~ & & \gamma \in \Gamma_{\delta},~\delta = (x_k,x_{k+1},\theta) \in \Delta.
\end{align}
Here, $\Delta$ and $\Gamma_{\delta}$ are as follows:
\begin{equation}
    \label{eq:setting}
    \begin{gathered}
        \Delta = \{(x_k,x_{k+1},\theta) \in \mathcal{X} \times \mathcal{X} \times \Theta~|~h(x_0,\theta) \geq 0 \}, \\
        \Gamma_{\delta} = \left\{\gamma \in \mathbb{R}~\bigg|~
        h(x_{k+1},\theta) \geq \gamma \left|h(x_k,\theta)\right|\right\}
    \end{gathered}
\end{equation}
Specifically, $\Delta$ is the set of transitions $x_k$ to $x_{k+1}$ for trajectories whose initial condition and parameter $(x_0,\theta)$ start in the $0$-superlevel set of the candidate barrier function $h$.  $\Gamma_{\delta}$ is the set of all $\gamma \in \mathbb{R}$ that satisfy an inequality similar to~\eqref{eq:desired_inequality} for the specific transition $x_k$ to $x_{k+1}$ encoded by $\delta$.  The discrepancy with~\eqref{eq:desired_inequality} is the absolute value over $h(x_k,\theta)$ - the reason for which will be elucidated in a lemma to follow.

Furthermore, for~\eqref{eq:uncertain_opt} to be an uncertain program, $\delta$ must be a random variable with an associated probability distribution $\pi$.  As such, we will define the probability of sampling any $\delta \triangleq (x_k,x_{k+1},\theta)$ as the probability of sampling an initial condition and parameter $(x_0,\theta) \in \mathcal{C}$ such that the corresponding closed-loop trajectory to~\eqref{eq:nominal_system} contains the transition encoded by $\delta$.  To formalize this distribution $\pi$, we can define the indicator function $\indicator_{\delta}$ for a given transition $\delta = (x_k,x_{k+1},\theta_d)$.  This function evaluates to $1$ for any initial condition and parameter pair $(x_0,\theta)$ such that the corresponding closed-loop trajectory to~\eqref{eq:nominal_system} contains the transition specified by $\delta$ within $K$ timesteps.
\begin{equation}
    \indicator_{\delta}(x_0,\theta) = \begin{cases}
        1 & \mbox{if~} \begin{cases}
        \theta = \theta_d,~\mathrm{and,~} \\
        \exists~k,k+1 \suchthat 0 \leq k,k+1 \leq K,\\
        x(t_k,\theta) = x_k,~x(t_{k+1},\theta) = x_{k+1},
        \end{cases} \\
        0 & \mbox{else.}
    \end{cases}
\end{equation}

Then, if we sample initial conditions and parameters $(x_0,\theta)$ with a uniform distribution over $\mathcal{C}$, we can implicitly define the probability distribution function $\pi$ for the random variable $\delta$.  In what follows, $s$ is a normalization constant to ensure the total probability integrates to $1$ and $\beta = (x_0,\theta)$:
\begin{equation}
    \label{eq:prob_distribution}
    \prob_{\pi(\delta)}[A \subset \Delta] = \int_{A} \int_{\mathcal{C}} \frac{\indicator_{\delta}(\beta)}{s} d\beta d\delta. 
\end{equation}
While it is currently unclear whether we can sample from this distribution $\pi$, we will show we can through a few examples in the section to follow.  Intuitively though, if we uniformly sample initial condition and parameter pairs $(x_0,\theta)$ and record the state trajectory at time-steps $t_k~\forall~k\leq K$, the corresponding transitions will be samples of $\delta$ from the proposed distribution $\pi$.

Under the assumption that we can take $N$ samples $\delta$ with corresponding probability distribution $\pi$, we can generate a scenario program for~\eqref{eq:uncertain_opt} as follows:
\begin{align}
        \gamma^*_N & = \argmax_{\gamma \in \mathbb{R}}~ & &\gamma, \label{eq:actual_scenario} \tag {BASE-RP-N} \\
        &~~\mathrm{subject~to}~ & & \gamma \in \Gamma_{\delta_i},~\forall~\delta_i\in\{\delta_i\}_{i=1}^N.
\end{align}
For a solution $\gamma^*_N$ to~\eqref{eq:actual_scenario} we can define an associated violation set $F(\gamma)$ and violation probability $V(\gamma)$.
\begin{equation}
    \label{eq:actual_violation}
        F(\gamma) = \{\delta \in \Delta~|~\gamma \not \in \Gamma_{\delta}\},\quad V(\gamma) = \prob_{\pi(\delta)}[F(\gamma)].
\end{equation}
Intuitively, $F(\Tilde \gamma)$ is the set of all transitions $\delta$ where the system decays to an unsafe behavior quicker than the minimum decay rate identified by $\Tilde \gamma$ through inequality~\eqref{eq:desired_inequality}.  $V(\Tilde \gamma)$ is the probability of sampling such transitions from $\pi$.  

\newidea{Description and Statement of Results:} This puts into place all required notation for our main results to follow.  Succinctly, we will prove that we can always solve~\eqref{eq:actual_scenario} for any $N$-sample set of transitions $\{\delta_i\}_{i=1}^N$, and that any solution $\gamma^*_N$ will be unique.  This statement and proof will be formalized through Lemma~\ref{lem:solvability}.  Then, through Lemma~\ref{lem:probability_equivalence} we will prove that the violation probability of our solution $V(\gamma^*_N)$ corresponds to the probability of sampling an unsafe trajectory when uniformly sampling initial conditions and parameters $(x_0,\theta) \in \mathcal{C}$.  Then, our main result will use both prior Lemmas and Theorem~\ref{thm:scenario_opt} to lower bound the probability of sampling safe trajectories over all possible initial conditions and parameters $(x_0,\theta) \in \mathcal{C}$ if the solution $\gamma^*_N$ to~\eqref{eq:actual_scenario} is positive.  Finally, Corollary~\ref{cor:counterexample} will extend Lemma~\ref{lem:probability_equivalence} and state that any negative solution to~\eqref{eq:actual_scenario} corresponds to a counterexample.  We will now state these results.

\begin{figure*}[t]
    \centering
    \includegraphics[width =\textwidth]{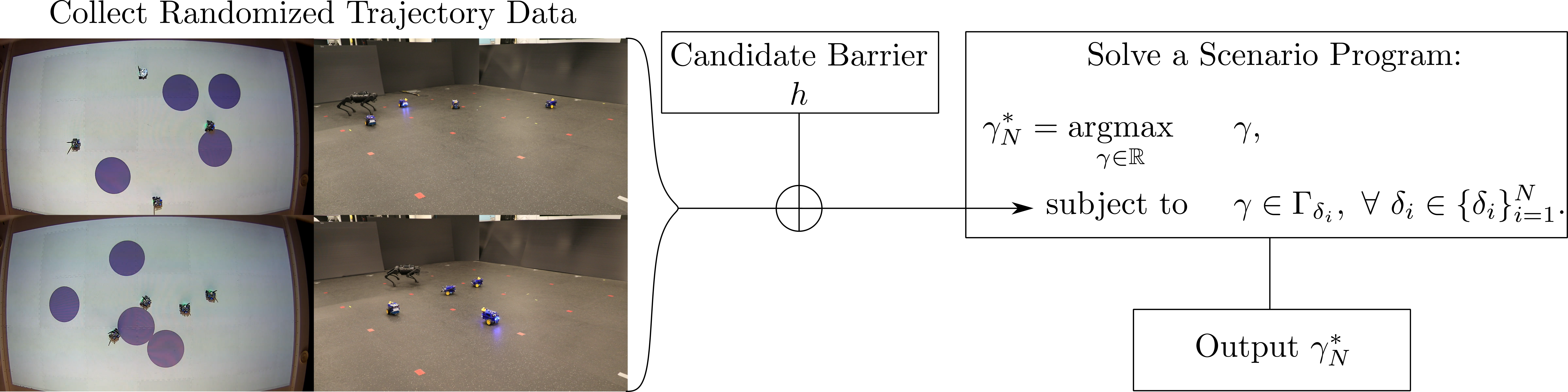}
    \caption{Shown above is an overview of the process detailed in the paper.  By recording state trajectory samples of randomly chosen trajectories and evaluating the transitions under a candidate barrier function $h$, we inform the constraints for a randomized linear program that identifies the minimum discrete-time decrement condition $\gamma^*_N$.  As per Theorem~\ref{thm:main_result} if $\gamma^*_N \geq 0$, then we prove that the corresponding system renders the candidate barrier function $h$ a control barrier function with high probability.  In other words, if $\gamma^*_N \geq 0$, then the system maintains positivity of $h$ with high probability.} \vspace{-0.2 in}
    \label{fig:overview}
\end{figure*}

\begin{lemma}
\label{lem:solvability}
Let Assumption~\ref{assump:barrier_existence} hold.  The scenario program~\eqref{eq:actual_scenario} is always solvable for any $N$-sample set of transitions $\{\delta_i\}_{i=1}^N$, and the solution $\gamma^*_N$ is unique.
\end{lemma}
\begin{lemma}
\label{lem:probability_equivalence}
The violation probability $V(\gamma^*_N)$ for a solution $\gamma^*_N$ to~\eqref{eq:actual_scenario} is equivalent to the probability of uniformly sampling over $\mathcal{C}$ initial conditions and parameters $(x_0,\theta)$ whose corresponding trajectory evolves within $K$ time-steps to a transition with a faster safety decay rate than $\gamma^*_N$, \textit{i.e.} with $x^\theta_k = x(t_k,\theta)$,
\begin{equation}
V(\gamma^*_N) = \prob_{\uniform[\mathcal{C}]}\left[(x_0,\theta)~\Bigg|
\begin{gathered}
\exists~k,k+1 \suchthat \\
0 \leq k,k+1 \leq K,~\mathrm{and}, \\
h(x^\theta_{k+1},\theta),\theta) < \gamma^*_N \left| h(x^\theta_k,\theta)\right|
\end{gathered} \right].
\end{equation}
\end{lemma}
\begin{theorem}
\label{thm:main_result}
Let Assumption~\ref{assump:barrier_existence} hold, let the scenario program~\eqref{eq:actual_scenario} be composed from an $N$-sample set of transitions $\{\delta_i\}_{i=1}^N$, and let $\epsilon \in [0,1]$.  If $\gamma^*_N \geq 0$, then the following statement is true, with $x^\theta_k = x(t_k,\theta)$:
\begin{equation}
    \begin{gathered}
        S(\gamma^*_N) \triangleq \prob_{\uniform[\mathcal{C}]}\left[(x_0,\theta)~|~h(x^\theta_k,\theta)\geq 0,~\forall~k = 0,1,\dots K\right], \\
        \prob^N_{\pi(\delta)}\left[S(\gamma^*_N) \geq 1-\epsilon \right] \geq 1 - (1-\epsilon)^N.
    \end{gathered}
\end{equation}
\end{theorem}
\begin{corollary}
\label{cor:counterexample}
Let Assumption~\ref{assump:barrier_existence} hold and let the scenario program~\eqref{eq:actual_scenario} be composed from an $N$-sample set of transitions $\{\delta_i\}_{i=1}^N$.  If $\gamma^*_N < 0$, then there exists a transition $\delta \in \{\delta_i\}_{i=1}^N$ that corresponds to a safety violation, \textit{i.e.} $ \exists~\delta \in \{\delta_i\}_{i=1}^N \suchthat h(x^\theta_{k+1}, \theta) < 0$ with $x^\theta_k = x(t_k,\theta)$ and $\delta = (x^\theta_k,x^\theta_{k+1}, \theta)$.
\end{corollary}
We can summarize Theorem~\ref{thm:scenario_opt} as follows.  If Assumption~\ref{assump:barrier_existence} holds, the corresponding scenario program is formed from $N$ samples, and $\gamma^*_N \geq 0$, then the probability of uniformly sampling over $\mathcal{C}$ initial conditions and parameters $(x_0,\theta)$ such that their corresponding trajectories remain safe for at least $K$ time-steps can be lower bounded with high probability.  We will now prove these results.

\subsection{Proofs of Main Results}
We will start first with the proof for Lemma~\ref{lem:solvability}.

\begin{proof}
To start, by definition of the scenario program~\eqref{eq:actual_scenario} and the constraint sets $\Gamma_{\delta}$ in equation~\eqref{eq:setting}, for any $N$-sized sample set of transitions $\{\delta_i\}_{i=1}^N$, the scenario program~\eqref{eq:actual_scenario} is a linear program maximizing the decision variable $\gamma$ subject to a set of upper bounds.  It is for this reason we required the absolute value over $h(x_k,\theta)$ in~\eqref{eq:setting}.  Without the absolute value, there could exist a case where the system evolves to a state where $h(x_k,\theta) < 0$, which has the potential of yielding an un-satisfiable set of constraints.  With the absolute value, proving solvability of the associated program requires ensuring that all upper bounds are strictly less than infinity.

This arises as evaluations of $h$ are restricted to lie within $[-m,M],~M,m \in \mathbb{R}_+$ by Assumption~\ref{assump:barrier_existence}.  As such, there is guaranteed to be at least one such upper bound $\Tilde \gamma < \infty$.  This guarantees a solution to the corresponding linear program with the solution guaranteed to be unique as it is a solution to a linear program. This neglects to consider those trajectories that eventually end up in a state $x_k$ such that $h(x_k,\theta) = 0$.  However, the set of all trajectories that land in the set $h(x_k,\theta) = 0$ for some $k = 0,1,\dots,K-1$ is a set of measure $0$ with respect to the probability distribution $\pi$.  Therefore, we can safely neglect such trajectories.
\end{proof}

Lemma~\ref{lem:solvability} effectively acts as a disclaimer permitting us to utilize the results of Theorem~\ref{thm:scenario_opt} to bound the violation probabilities of results to our scenario program~\eqref{eq:actual_scenario}.  This will be useful, for as stated in Lemma~\ref{lem:probability_equivalence}, this violation probability is equivalent to the probability of sampling marginally "more unsafe" trajectories.  The proof for Lemma~\ref{lem:probability_equivalence} will follow.

\begin{proof}
We will start with the definition of the violation probability for our optimal solution $V(\gamma^*_N)$ as per~\eqref{eq:actual_violation}.
\begin{equation}
    V(\gamma^*_N) = \prob_{\pi(\delta)}[\delta \in \Delta~|~\gamma^*_N \not \in \Gamma_{\delta}].
\end{equation}
Here, $\Gamma_{\delta}$ is defined in equation~\eqref{eq:setting}.  For this proof, it will be useful to define the following indicator function:
\begin{equation}
    \neg \indicator_{\Gamma_{\delta}}(\gamma) = \begin{cases}
    1 & \mbox{if~} \gamma \not \in \Gamma_{\delta}, \\
    0 & \mbox{else}.
    \end{cases}
\end{equation}
Then we can rewrite the violation probability as follows:
\begin{equation}
    V(\gamma^*_N) = \int_{\Delta} \pi(\delta) \neg \indicator_{\Gamma_{\delta}}(\gamma^*_N) d \delta.
\end{equation}
By definition of our probability distribution $\pi$ in equation~\eqref{eq:prob_distribution}, we can rewrite the above equation with $\beta = (x_0,\theta)$:
\begin{equation}
    V(\gamma^*_N) = \int_{\Delta} \int_{\mathcal{C}}\frac{\indicator_{\delta}(\beta) \neg \indicator_{\Gamma_{\delta}}(\gamma^*_N)}{s} d\beta d \delta.
\end{equation}
However, the interior integrand only evaluates to $1$ for those initial condition and parameter pairs $(x_0,\theta)$ such that there exist time-steps $k,k+1$ where $0 \leq k, k+1 \leq K$ and $h(x^\theta_{k+1},\theta) < \gamma^*_N \left| h(x^\theta_k,\theta)\right|$.  Here, $x^\theta_k = x(t_k,\theta)$.  As such, the above probability corresponds to the probability of sampling such initial condition and parameter pairs from the uniform distribution over $\mathcal{C}$.  This completes the proof.
\end{proof}

Lemma~\ref{lem:probability_equivalence} effectively states that the violation probability for our scenario program~\eqref{eq:actual_scenario} corresponds to picking those trajectories that are marginally "more unsafe" than the worst-case sampled trajectory.  In other words, one minus the violation probability then corresponds to the probability of picking those trajectories that are at least as safe as the worst-case sampled trajectory.  This notion is formalized in Theorem~\ref{thm:main_result} the proof for which will follow.

\begin{proof}
With the assumptions behind Theorem~\ref{thm:main_result}, we know that Lemma~\ref{lem:probability_equivalence} holds.  This let's us define the success probability $S(\gamma^*_N) = 1-V(\gamma^*_N)$.  Mathematically this success probability is defined as follows with $x^\theta_k = x(k,\theta)$:
\begin{equation}
S(\gamma^*_N) = \prob_{\uniform[\mathcal{C}]}\left[(x_0,\theta)~\Bigg|
\begin{gathered}
\forall~k = 0,1,\dots,K-1 \\
h(x^\theta_{k+1},\theta) \geq \gamma^*_N \left|h(x^\theta_k,\theta)\right|
\end{gathered} \right].
\end{equation}
For all such sampled trajectories, $h(x^\theta_0,\theta) \geq 0$ as the initial condition and parameter $(x_0,\theta)$ are sampled uniformly over the $0$-superlevel set of $h$ $\mathcal{C}$.  As a result, in light of Theorem~\ref{thm:disc_cbf} we can rewrite the condition for the success probability:
\begin{equation}
S(\gamma^*_N) = \prob_{\uniform[\mathcal{C}]}\left[(x_0,\theta)~|
h\left(x^\theta_k,\theta\right) \geq 0,~\forall~k=0,1,\dots,K \right].
\end{equation}
This satisfies the first equality for Theorem~\ref{thm:main_result}.

\begin{figure*}[t]
    \centering
    \includegraphics[width = 0.99\textwidth]{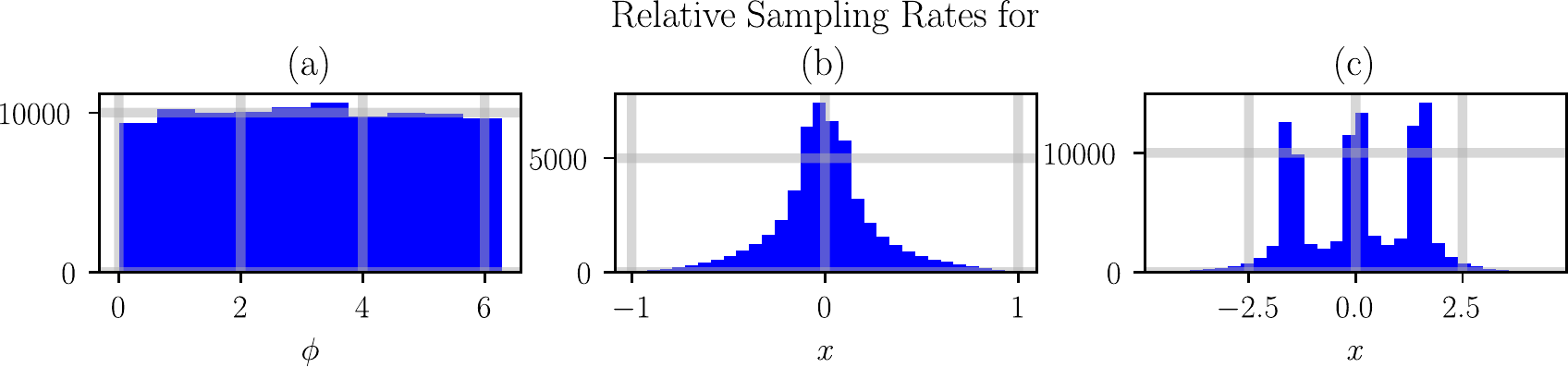}
    \caption{Relative sampling rates of the initial state $x_k$ for transitions recorded by our uncertain parameter $\delta = (x_k,x_{k+1},\theta)$.  The systems for which these transitions are sampled are given in equations~\eqref{eq:system_a}-\eqref{eq:system_c}.  In each case, however, our proposed sampling method does produce numerical estimates for the distribution of the initial states $x_k$ that align with our expectations based on definition of our proposed distribution $\pi$ in equation~\eqref{eq:prob_distribution}.}
    \vspace{-0.4 cm}
    \label{fig:sampling}
\end{figure*}

For the inequality in Theorem~\ref{thm:main_result}, we first note that Lemma~\ref{lem:solvability} permits us to use the results of Theorem~\ref{thm:scenario_opt}.  This lets us upper bound the violation probability to high confidence.  Note that for our problem $d=1$ which lets us simplify the right hand side of the inequality in Theorem~\ref{thm:scenario_opt}.  Specifically, for some $\epsilon \in [0,1]$,
\begin{equation}
    \prob^N_{\pi(\delta)}\left[V(\gamma^*_N) \leq \epsilon\right] \geq 1 - (1-\epsilon)^N.
\end{equation}
Then the final result holds due to definition of the success probability $S(\gamma^*_N)$.
\begin{equation}
     \vspace{-0.675 cm}
    \prob^N_{\pi(\delta)}\left[S(\gamma^*_N) \geq 1-\epsilon\right] \geq 1 - (1-\epsilon)^N.
\end{equation} \vspace{0.075 in}
\end{proof}

This ends the proof for our main result - that if the solution to our randomized linear program~\eqref{eq:actual_scenario} is positive, \textit{i.e.} $\gamma^*_N \geq 0$, then with high probability the system maintains positivity of the candidate barrier function $h$ for at least $K$ time-steps.  What if $\gamma^*_N < 0$, however?  Corollary~\ref{cor:counterexample} indicates that such a scenario corresponds to a counterexample and its proof will follow.

\begin{proof}
The proof for this corollary stems primarily from the definition of the constraint spaces $\Gamma_{\delta}$ in~\eqref{eq:setting}.  Specifically, by Lemma~\ref{lem:solvability}, we know a solution to~\eqref{eq:actual_scenario} must exist for any sample set of transitions $\{\delta_i\}_{i=1}^N$.  As mentioned in the proof for Lemma~\ref{lem:solvability} this is primarily due to the fact that for any set of samples,~\eqref{eq:actual_opt} is a linear program maximizing a scalar decision variable $\gamma$ subject to a set of upper bounds $b_i$.  If the solution $\gamma^*_N < 0$ this implies that at least one upper bound $b_i < 0$.  Based on definition of the constraint space $\Gamma_{\delta}$ then, this implies that
\begin{equation}
    \exists~\delta = (x^\theta_k, x^\theta_{k+1}, \theta) \in \{\delta_i\}_{i=1}^N \suchthat \frac{h\left(x^\theta_{k+1},\theta\right)}{\left|h\left(x^\theta_{k},\theta\right)\right|} < 0.
\end{equation}
As the denominator for the associated fraction is always positive, this implies that $\exists~\delta \in \{\delta_i\}_{i=1}^N$ such that $h(x^\theta_{k+1},\theta) < 0$, concluding the proof.
\end{proof}

This concludes all proofs for our results.  As mentioned, however, these results hinge on the capacity to take samples of the random variable $\delta$ with distribution $\pi$.  The following section will show a few examples indicating that we can sample from our proposed distribution.

\subsection{Sampling from our Proposed Distribution}
As mentioned earlier, it is unclear whether we can sample from our proposed distribution $\pi$ as defined in equation~\eqref{eq:prob_distribution}.  However, we offered a method to take samples from this distribution.  Our method first uniformly randomly samples the initial condition and parameter pair $(x_0,\theta)$ from $\mathcal{C}$ and records the resulting state trajectory at time-steps $t_k,~\forall~k=0,1,\dots,K$. This section will show a few numerical examples indicating that this method does produce samples of $\delta$ distributed by $\pi$.

We will first provide three simple systems to act as a replacement for the continuous system we are trying to verify.  The first will be a system that continuously oscillates around the perimeter of a circle with radius $r=1$, the other will be a system that exponentially decays to $0$, and the third will be a system that exponentially decays to a parameterized point $\theta \in [-1.5,0,1.5]$.  We will not mention their ODEs for motion and only mention their solutions.
\begin{align}
    x & = [1,\phi]^T,~& & \hspace{-0.25 in} x(t) = [1, \phi_0 + 0.1t], \label{eq:system_a} \tag{a} \\
    x & = [x],~& & \hspace{-0.25 in} x(t) = [e^{-0.5 t}]. \label{eq:system_b} \tag{b} \\
    x & = [x],~\theta \in [-1.5,0,1.5],~& & \hspace{-0.25 in} x(t) = [e^{-3t}+\theta]. \label{eq:system_c} \tag{c}
\end{align}
For sampling purposes then, the respective sample spaces per system are as follows:
\begin{equation}
    \label{eq:example_spaces}
    \begin{aligned}
    \mathrm{for~\eqref{eq:system_a}}~\mathcal{C} & = \mathcal{X} = [0,2\pi], \\
    \mathrm{for~\eqref{eq:system_b}}~\mathcal{C} & = \mathcal{X} = [-1,1], \\
    \mathrm{for~\eqref{eq:system_c}}~\mathcal{C} & = \mathcal{X} \times \Theta = [-3,3] \times [-1.5,0,1.5].
    \end{aligned}
\end{equation}

Per our method then, we will take $500$ trials of each system, forward simulating each system $K = 200$ steps per trial with $\Delta t = 0.05$.  For~\eqref{eq:system_c} we will increase the trial number to $1000$ trials as we are parameterizing the system via $\theta$ as well.  To generate these trials, we will uniformly randomly sample an initial condition (and parameter $\theta$ for~\eqref{eq:system_c}) with the spaces shown in equation~\eqref{eq:example_spaces}.  For data portrayal purposes, we will show in Figure~\ref{fig:sampling} the initial states $x_k$ for each sampled transition $\delta = (x_k,x_{k+1},\theta)$, as they suffice to showcase our method's ability to capture the intended distribution $\pi$.

For system~\eqref{eq:system_a} if we uniformly sample initial conditions and forward simulate the same number of steps for each initial condition, we expect the initial states $x_k$ for each transition to follow a uniform distribution.  This is indeed the case as seen in the left figure in Figure~\ref{fig:sampling}.  For system~\eqref{eq:system_b}, we expect the initial states $x_k$ to be localized to and symmetric about $0$, with an exponentially higher rate of samples closer to $0$ than farther out.  This harmonizes with the numerical results shown in the center figure in Figure~\ref{fig:sampling}.  Finally, for system~\eqref{eq:system_c}, we expect a response similar to that for system~\eqref{eq:system_b} but for three different "peaks" centered on the choice of $\theta \in [-1.5,1,1.5]$.  As seen in the right figure in Figure~\ref{fig:sampling}, this is indeed the case. Therefore, these examples show that we can sample from our proposed distribution $\pi$ with the method we describe, and will now use this method to verify a system simulator and its hardware counterpart. 

\section{Examples}
In this section, we will verify or find counterexamples for systems with pre-existing controllers.  Furthermore, we will provide numerical evidence that the stated inequality in Theorem~\ref{thm:main_result} is true. We will start with verifying the Robotarium simulator~\cite{robotarium} which will provide numerical results supporting the results of Lemma~\ref{lem:probability_equivalence} and Theorem~\ref{thm:main_result}.
\subsection{Verifying the Robotarium Simulator}
\label{sec:sim_results}
The robots in the robotarium are modeled via unicycle dynamics which are as follows:
\begin{equation}
    \begin{gathered}
    x = \begin{bmatrix}
    x,\\
    y, \\
    \theta
    \end{bmatrix},~\dot x = 
    \begin{bmatrix}
    v \cos(\theta), \\
    v \sin(\theta), \\
    \omega,
    \end{bmatrix},~u = [v, \omega]^T, \\
    \mathcal{X} = [-1.2,1.2] \times [-0.6,0.6] \times [0,2 \pi],~P = [I_2,~\mathbf{0}_{2x1}]
    \end{gathered}
\end{equation}
Each robot in the robotarium has a Lyapunov-based controller that drives it from its current position to a desired orientation $x_d$ in its state space $\mathcal{X}$.  When multiple robotarium robots are asked to ambulate in the same, confined space, their control inputs are filtered in a barrier-based quadratic program to ensure that the robots never collide~\cite{ames2016control}.  As such, given a nominal radius $r_s$ that the robots are to maintain, a candidate barrier function $h$ would be
\begin{equation}
    h(x^1,x^2,\dots,x^{N_R}) = \min_{i \neq j,~i,j \in [1,2,\dots,N_R]}~\|P(x^i - x^j)\| - r_s.
\end{equation}
Concatenating all the state vectors of the $N_R$ robots in the robotarium in $\mathbf{x}^T = [x^{1T}, x^{2T}, \dots x^{NT}]$, we get the following candidate barrier function required of Assumption~\ref{assump:barrier_existence}:
\begin{equation}
    \label{eq:cand_barrier_robotarium}
    h(\mathbf{x}) = \min_{i \neq j,~i,j \in [1,2,\dots,N_R]}~\|P(x^i - x^j)\| - r_s.
\end{equation}
This results in the following verification problem.  For the closed-loop robotarium simulator, devise a method to determine whether $h(\mathbf{x}(t_k,\mathbf{x}_d)) \geq 0~\forall~k \leq K=100\in\mathbb{Z}_+$ and $(\mathbf{x}_0,\mathbf{x}_d) \in \mathcal{C}$.  Here, $h$ is as per~\eqref{eq:cand_barrier_robotarium}, our parameter $\theta^T = [x^{1T}_d,x^{2T}_d,\dots x^{N_R}_d] \in \Theta = \mathcal{X}^{N_R}$, $x^j_d$ is the desired pose for robot $j$, and $N_R = 3$.

\spacing
\newidea{Numerical Results:} To determine the safety of the robotarium simulator, we sampled $N_0 = 100$ initial conditions and desired poses $(\mathbf{x}_0,\mathbf{x}_d)$ from the uniform distribution over the $0$-superlevel set of the candidate barrier function $h$ $\uniform[\mathcal{C}]$.  Then, we simulated each closed-loop trajectory for $K = 100$ time-steps with $\Delta t = 0.03$ and recorded all transitions $\delta = (\mathbf{x}_k, \mathbf{x}_{k+1},\mathbf{x}_d)$.  Then, we calculated the minimum decay constant $\gamma^*_N$ as per~\eqref{eq:actual_scenario} for the $N = 10000$ transition samples taken, resulting in $\gamma^*_N \approx 0.953 \geq 0$.  

As per Lemma~\ref{lem:probability_equivalence} and Theorem~\ref{thm:main_result} then, the probability of sampling a violating initial condition and goal $(\mathbf{x}_0,\mathbf{x}_d)$ from $\uniform[\mathcal{C}]$ can be upper bounded by some $\epsilon \in [0,1]$.  More accurately, Theorem~\ref{thm:main_result} states that the probability that the system maintains positivity of $h$ for at least $K=100$ time-steps can be lower bounded as follows:
\begin{equation}
    \prob^N_{\pi(\delta)}\left[S(\gamma^*_N) \geq 1-\epsilon \right] \geq 1 - (1-\epsilon)^N.
\end{equation}
To determine the high probability lower bound $1-\epsilon$, we set the right-hand side of the outer probability to be $= 1-10^{-6}$ and calculate the corresponding violation probability upper bound $\epsilon$ that satisfies this inequality with $N = 10000$, the number of samples taken.  This results in an upper bound $\epsilon = 0.0014$.  In other words, according to the results of Theorem~\ref{thm:main_result}, the probability of sampling an initial condition and goal pair $(\mathbf{x}_0,\mathbf{x}_d)$ from $\uniform[\mathcal{C}]$ such that the corresponding trajectory does not maintain positivity of $h$ for at least $K=100$ time-steps should be lower than $\epsilon = 0.0014$ with minimum probability $1-10^{-6}$.

\begin{figure*}[t]
    \centering
    \includegraphics{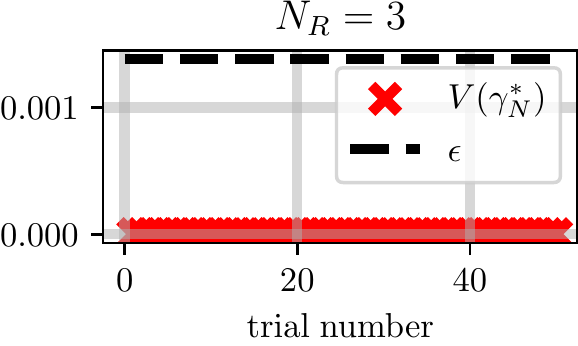}
    \includegraphics{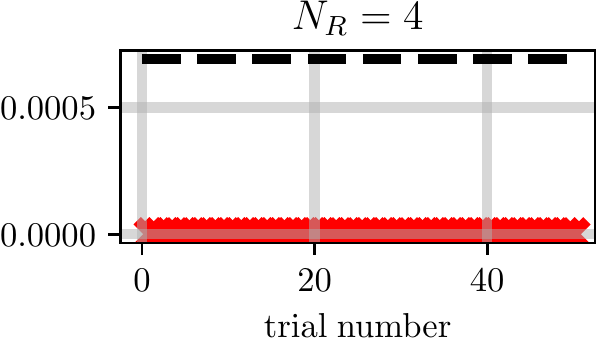}
    \includegraphics{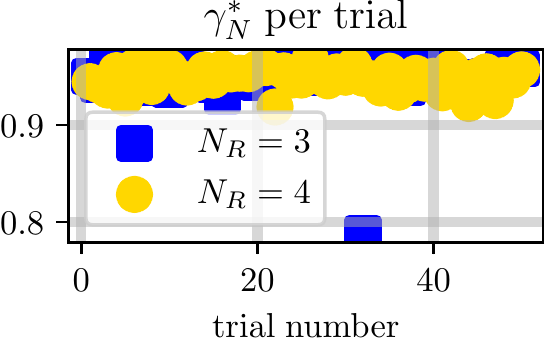}
    \caption{Compilation of results for each of the $50$ trial runs of verifying the robotarium simulator with $N_R=3,4$ robots.  A complete explanation of the results is in Section~\ref{sec:sim_results}.  Notably, however, over all $100$ separate trials performed, the true violation probability of our calculated decay constant $\gamma^*_N$ is always less than our theorized upper bound $\epsilon$.  Just to note, while it may seem like the true violation probability is equal to $0$ in all cases, it is not.  The variance in the true violation probability is on the order of $10^{-5}$ and is just not visible at the scale shown.}
    \vspace{-0.4 cm}
    \label{fig:sim_info}
\end{figure*}

To verify this last statement, we sampled $N_0 = 50000$ initial condition and goal pairs $(\mathbf{x}_0,\mathbf{x}_d)$ from $\uniform[\mathcal{C}]$, simulated each trajectory for $K=100$ timesteps, and recorded the minimum barrier value $\ell$ for the sampled trajectory.  We used the fraction of trajectories with minimum barrier value $\ell < 0$ to approximate the probability of sampling a trajectory that does not maintain positivity of $h$ for at least $100$ time-steps.  The complete information for this verification process and validation of our verification method can be found in Table~\ref{table:sim_verification}.  As shown in this information, the true violation probability estimate is indeed lower than our calculated upper bound $\epsilon$.  This serves as numerical validation of our verification method, at least with respect to the Robotarium simulator.

\begin{table}[t]
    \centering
    \caption{Robotarium Simulator Verification Data}
    \label{table:sim_verification}
    \begin{tabular}{|c|c|c|c|c|c|c|}
        \hline
        $N_0$ & $K$ & $N$ & $\gamma^*_N$ & $\epsilon$ & $V(\gamma^*_N)$ & $S(\gamma^*_N)$  \\
        \hline
        100 & 100 & 10000 & 0.953 & 0.0014 & $\approx 0$ & $\approx 1$ \\
        \hline
    \end{tabular}
\end{table}

\spacing
\newidea{Verifying our Scenario Approach:} The aforementioned results provide numerical evidence supporting the results of Theorem~\ref{thm:main_result} and Lemma~\ref{lem:probability_equivalence}.  However, this does not show repeatability of our results.  Specifically, we state via use of the scenario approach that the violation probability of our calculated solution $\gamma^*_N$ can be upper bounded with high probability with respect to the distribution $\pi$ by which transition samples $\delta$ are drawn, \textit{i.e.}
\begin{equation}
    \prob^N_{\pi(\delta)}\left[ V(\gamma^*_N) \leq \epsilon \right] \geq 1-(1-\epsilon)^N.
\end{equation}

To show the above statement holds, we performed the same verification procedure as prior $50$ separate times and recorded the calculated minimum safety decay rate $\gamma^*_N$ each time.  To show that our results are also system independent, we performed the same procedure with a four-agent system as well and extended the maximum number of time-steps to $K=200$ in the four-agent case.  Then, for each verification attempt, we calculated the true violation probability of the our solution $\gamma^*_N$ by uniformly sampling another $N_v = 1000$ initial condition and goal pairs and simulating the corresponding trajectories for the appropriate number of time-steps as well - $K=100$ for the three-agent case $N_R = 3$ and $K=200$ for the four-agent case $N_R = 4$.  Then, we recorded the fraction of transitions $\delta$ that required decay constants $\gamma < \gamma^*_N$ as an estimate of the true violation probability of our solution $\gamma^*_N$.  Finally, we calculated the high probability upper bound $\epsilon$ to our violation probability $V(\gamma^*_N)$ by setting $1-(1-\epsilon)^N = 1-10^{-6}$ and calculated the $\epsilon$ satisfying this condition.  For $N_R = 3$ $\epsilon = 0.0014$ and for $N_R=4$ $\epsilon = 0.0007$.  This discrepancy in $\epsilon$ arises as we changed the maximum simulation time-step from $K= 100$ to $K=200$ from the three to four-agent case respectively.

All information for this procedure can be found in Figure~\ref{fig:sim_info}.  Notably, over all $100$ trials performed, the calculated minimum decay constant $\gamma^*_N \geq 0$.  This acts as further numerical support for the results of Lemma~\ref{lem:probability_equivalence} and Theorem~\ref{thm:main_result}.  Specifically, our prior results indicated that with high probability, uniformly sampled trajectories should consistently render positive the candidate barrier function $h$ for at least $100$ time-steps.  Indeed, over all $100$ trials performed wherein each trial $100$ trajectories were sampled, the candidate barrier function $h$ stayed positive for at least $100$ time-steps.  Furthermore, for all $100$ trials, the true violation probability $V(\gamma^*_N) < \epsilon$ - our calculated upper bound.  This shows the repeatability and accuracy of our verification attempts at least with respect to the Robotarium simulator.  The next section aims to show similar results on hardware systems as well.

\subsection{Hardware Verification with Limited Data}
\label{sec:hardware_verification}
In this section, we will verify two hardware systems, the Robotarium, and the Unitree A1 Quadruped steered by a variant of the controller presented in~\cite{molnar2021model}.  The robotarium experiments will provide further numerical validation of our scenario approach to verification.  Namely, we will calculate a minimum safety decay constant $\gamma^*_N$ based on $100$ randomly sampled hardware trajectories and show the true violation probability of this solution $V(\gamma^*_N)$ is indeed upper bounded as stated via our approach.  The quadruped examples show how we can also make this verification statement with limited data independent of the system-to-be-verified.

\spacing
\newidea{Robotarium:} The setting for the hardware version of the robotarium is the same as mentioned in the simulation section prior for the three-agent case.  Our goal will be to verify the same probabilistic statement as prior: $\prob^N_{\pi(\delta)}\left[ V(\gamma^*_N) \leq \epsilon \right] \geq 1-(1-\epsilon)^N.$  Here, $\pi$ is now the unknown distribution from which transitions $\delta$ of the hardware system are drawn.  To generate our minimum safety decay rate $\gamma^*_N$, we sampled $N_0=100$ initial condition and goal pairs, recorded the resulting trajectory for $K=100$ time-steps, and calculated $\gamma^*_N$ as per~\eqref{eq:actual_scenario} - this is the same procedure as we followed in the simulation case.  To verify our corresponding probabilistic result, we also sampled $N_v = 400$ initial condition and goal pairs, recorded the resulting trajectory for $K=100$ time-steps, and recorded the fraction $\ell$ of transitions $\delta$ requiring a safety decay rate $\gamma < \gamma^*_N$.  We used this fraction $\ell$ as an approximation of the true violation probability.  We also calculated the violation probability upper bound $\epsilon$ by setting the right-hand side of the earlier probability statement to $1-10^{-6}$ as prior.  This yields an upper bound $\epsilon = 0.0014$. All this information can be found in Table~\ref{table:hardware_verification}.

\begin{figure}[t]
    \centering
    \hspace{-0.2 cm} \includegraphics[width = 0.48\textwidth]{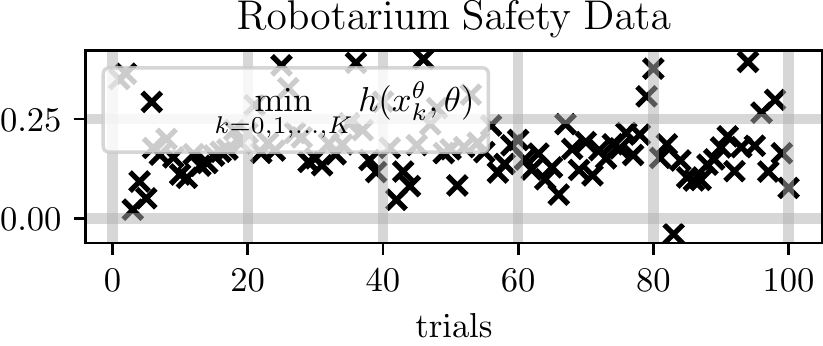}
    \caption{Minimum barrier value for each of the $100$ trials taken for the verification process of the robotarium hardware as stated in subsection~\ref{sec:hardware_verification}.  Notice that in one trial the system fails to keep positive the candidate barrier $h$ indicating a counterexample.}\vspace{-0.2 in}
    \label{fig:robotarium_hardware}
\end{figure}

Notice that in this case, the procedure identified a counterexample as $\gamma^*_N < 0$, and as per Corollary~\ref{cor:counterexample}, there should exist a transition $\delta$ in the sampled set whereby the system evolves to make negative the candidate barrier $h$.  This is indeed the case as shown in Figure~\ref{fig:robotarium_hardware}, as for one of the trials, the minimum value of the barrier function $h$ was negative.  That being said, over the other $400$ trajectories sampled, none exhibited a transition corresponding to a safety decay rate $\gamma < \gamma^*_N$.  As a result, we estimate that the true violation probability $V(\gamma^*_N) < \epsilon$ our calculated upper bound - a statement which we expected to hold with high probability, and it indeed does in this case.

\begin{table}[t]
    \centering
    \caption{Robotarium Hardware Verification Data}
    \label{table:hardware_verification}
    \begin{tabular}{|c|c|c|c|c|c|c|}
        \hline
        $N_0$ & $K$ & $N$ & $\gamma^*_N$ & $\epsilon$ & $V(\gamma^*_N)$ & $S(\gamma^*_N)$  \\
        \hline
        100 & 100 & 10000 & -3.057 & 0.0014 & $\approx 0$ & $\approx 1$ \\
        \hline
    \end{tabular}
\end{table}

\spacing
\newidea{Quadruped:} In a similar fashion as prior, we hope to verify the quadruped's ability to maintain positivity of a simple, 2-norm barrier function:
\begin{equation}
    h(x,\theta_1,\theta_2,\theta_3,\theta_4) = \min_{i = 1,2,3,4}\|x - \theta_i\| - 0.35.
\end{equation}
Here, we assume that our "state" $x$ is a projection of the true quadruped state onto the $x-y$ plane and constrained to lie within the state space $\mathcal{X} = [-1,2]^2$.  Also, $\theta_i$ is the location in the $x-y$ plane of one of the $4$ stationary obstacles.  As such, our total parameter vector $\theta \in \Theta = [-1,2]^8$.  Then, our verification problem is very similar to that which we had for the robotarium - see if the quadruped coupled with the controller in~\cite{molnar2021model} can keep positive this candidate barrier function $h$ for at least $K = 150$ time-steps with $\Delta t = 0.1$.

To make a probabilistic verification statement in this vein, we sampled $N_0 = 50$ initial conditions and parameter pairs $(x_0,\theta)$ from the uniform distribution over the $0$-superlevel set for the candidate barrier function $h$ $\uniform[\mathcal{C}]$.  We recorded the resulting trajectories for $K=150$ time-steps and recorded all transitions.  To be more specific about this process, we recorded state data at $1000$ Hz and recorded as the state $x_k$ the sampled state whose timestamp was nearest to the desired time $k\Delta t$.  This yielded $N = 7500$ transition samples and a calculated safety decay constant $\gamma^*_N  = 0.3931$.  Furthermore, we expect the violation probability for our solution $V(\gamma^*_N)$ to be upper bounded by $\epsilon = 0.0019$ with minimum probability $1-10^{-6}$.  This entire procedure required $12.5$ minutes of system data.  This is why we claim that we can verify hardware systems to high minimum probability with limited data and are confident in the validity of our approach based on the repeatability and validity analyses carried out earlier.

\section{Conclusion}
In this paper, we detail a randomized verification method for safety-critical systems with limited data via a scenario approach based on barrier functions.  We showed that by uniformly sampling initial conditions and parameters and recording the resulting state trajectory, we can determine the constraints for a randomized linear program designed to identify the minimum safety decay constant $\gamma$ required by the system in its attempt to maintain the positivity of a candidate barrier function $h$.  Given a sufficient number of trajectory samples, the positivity or lack thereof of this constant $\gamma$ provides us either a probabilistic verification statement or counterexample, respectively.  Finally, we showed that this procedure works across multiple systems, both simulated and real ones, and verified our probabilistic verification statements by taking copious samples of the same systems and showing our results holds.  As future work, we hope to extend our analysis to the case where the system dynamics are corrupted by additive noise and identify a similar approach for continuous barrier functions as well.

\bibliographystyle{IEEEtran}
\bibliography{IEEEabrv,bib_works}

\end{document}